\documentclass[twoside]{article}
\usepackage{fleqn,espcrc2}
\usepackage{epsfig}

\newcommand{\AmS}{{\protect\the\textfont2
  A\kern-.1667em\lower.5ex\hbox{M}\kern-.125emS}}

\title{CP parameters of the $B$ systems from Tevatron}

\author{S.~Burdin\address{University of Liverpool}%
       \  on behalf of CDF and D\O\   collaborations}
       
\begin{document}

\begin{abstract}
Recent results on CP parameters of the $B$ systems
obtained by the CDF and D\O\ collaborations using the data samples 
collected at 
the Tevatron Collider in the period 2002~--~2007 were presented
at the QCD 2008 conference (Montpellier, France). These results include 
measurements of the mixing phase $\phi_{s}^{J/\psi\phi}$, 
decay width difference $\Delta \Gamma_{s}$, and CP violation
parameters in the $B_{s}$ and $B_{u}$ decays.
\end{abstract}

\maketitle

\section{Introduction}
 The CDF and D\O\ 
experiments have successfully collected data since start of the Run II at the Tevatron
Collider in 2001. The presented results correspond to an integrated luminosity of up to 2.8~fb$^{-1}$ at each experiment.
The large $B$-meson production cross-section allows enough data to be collected to
study the tiny effects of 
CP-violation in $B$ systems. In particular, the results described in 
this paper cover measurements of  the mixing phase $\phi_{s}^{J/\psi\phi}$, 
decay width 
difference $\Delta\Gamma_{s}$, semileptonic asymmetry $a^{sl}_{s}$ in the $B_{s}$ system and direct CP violation in the $B_{s}$ and 
$B_{u}$ systems.

  A theoretical overview of $B_s - \bar{B}_s$ mixing is given in~\cite{dunietz,lenz}.
  Flavour eigenstates of $B_{s}$ meson propagate according to the Schr\"{o}dinger equation:
\begin{equation}
i\frac{d}{dt}
\left(
\begin{array}{c}
|B^{0}_{s}(t)\rangle \\
|\bar{B}^{0}_{s}(t)\rangle
\end{array}
\right) 
=
\left(
M_{s}-i\frac{\Gamma_{s}}{2}
\right)
\left(
\begin{array}{c}
|B^{0}_{s}(t)\rangle \\
|\bar{B}^{0}_{s}(t)\rangle
\end{array}
\right)
\label{eq:shreq}
\end{equation}

  The physical eigenstates $|B_{H}\rangle$ and $|B_{L}\rangle$ have different masses and lifetimes:
\begin{equation}
\Delta M_{s}=M^{s}_{H}-M^{s}_{L} = 2|M_{12}^{s}|,
\label{eq:deltams}
\end{equation}
\begin{equation}
\Delta \Gamma_{s}=\Gamma^{s}_{L}-\Gamma^{s}_{H} = 2|\Gamma_{12}^{s}| cos\phi_{s},
\label{eq:deltags}
\end{equation}
where 
\begin{equation}
\phi_{s}=arg(-M^{s}_{12}/\Gamma^{s}_{12})
\label{eq:phis}
\end{equation}
is the CP phase. 
The CP phase for the $B_{s}$ system is very small in the Standard Model (SM): 
$\phi_{s}^{SM}\approx 0.004$. The review~\cite{lenz} predicts 
that the CP phase $\phi_{s}$ is the most sensitive to the new physics effects. 
The semileptonic asymmetry defined as 
\begin{equation}
a^{s}_{sl}=\frac{N(\bar{B}_{s}\to f) - N(B_{s}\to \bar{f})}{N(\bar{B}_{s}\to f) + N(B_{s}\to \bar{f})},
\label{eq:asl_def}
\end{equation}
where $f$ corresponds to direct $B_{s}$ decays $B_{s}\to f$ (e.g. $D^{-}_{s}l^{+}\nu_{l}$), is 
related to the CP phase:
\begin{equation}
a^{s}_{sl}=\frac{\Delta\Gamma_{s}}{\Delta M_{s}}\tan \phi_{s}.
\label{eq:asl_phis}
\end{equation}
Measurements of all four parameters in the equation~\ref{eq:asl_phis} allow testing of the SM 
predictions with high sensitivity. The mixing phase $\beta_{s}$ with the SM prediction
$\beta_{s}^{SM}= arg(-(V_{ts}V_{tb}^{*})/(V_{cs}V_{cb}^{*}))\approx 0.02$ 
measured in decays $B_{s}\to J/\psi \phi$ is also related to the phase $\phi_s$ (see note in~\cite{lenz2}). 
In paper~\cite{d0_jpsiphi} the notation $\phi_{s}=-2\beta_{s}$ is used, but to distinguish 
this phase from the CP phase $\phi_{s}$ defined above in equation~\ref{eq:phis}
 the notation $\phi_{s}^{J/\psi\phi}=-2\beta_{s}^{J/\psi\phi}$
as in~\cite{hfag} is preferable.
The new physics contribution could be described by the phase $\phi_{s}^{NP}$ which modifies the original 
mixing phases in the following way: $\phi_{s}=\phi^{SM}_{s}+\phi_{s}^{NP}$ and 
$\beta_{s}=\beta^{SM}_{s}-\phi_{s}^{NP}/2$. 

\section{$\phi_{s}^{J/\psi\phi}$ from $B_{s}\to J/\psi \phi$~\cite{d0_jpsiphi,cdf_jpsiphi}}
\begin{figure}[htb]
\vspace{9pt}
\epsfig{figure=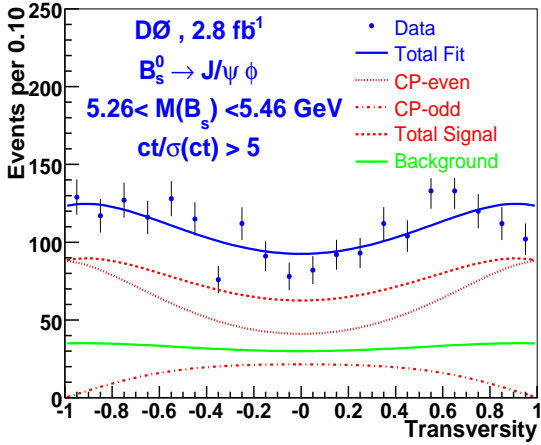,width=0.45\textwidth,height=0.3\textheight}
\caption{The transversity distribution for the signal-enhanced $J/\psi\phi$ subsample: 
“non-prompt” and signal mass region. The curves show: the signal contribution,
dotted (red); the background, light solid (green); and total, dark solid (blue).}
\label{fig:d0_jpsitrans}
\end{figure}

\begin{figure}[htb]
\vspace{9pt}
\epsfig{figure=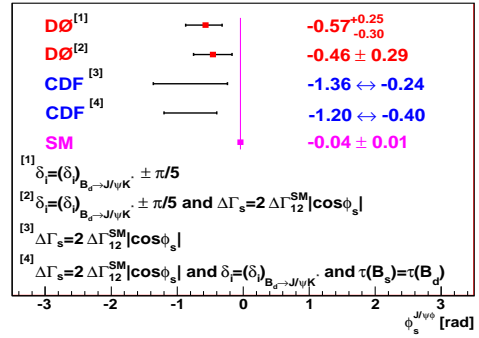,width=0.48\textwidth,height=0.25\textheight}
\caption{Comparison of the $\phi_{s}^{J/\psi\phi}$ 
results from CDF and D\O\   with the SM prediction.}
\label{fig:phis_comb}
\end{figure}

   The system $J/\psi + \phi$ from $B_{s}$ decay represents an admixture of CP-even and
CP-odd states. Angular analysis could be used to separate the
CP components (see figure~\ref{fig:d0_jpsitrans}) and measure the lifetime of each component and 
phase $\phi_{s}^{J/\psi\phi}$~\cite{dighe}.  Additional information on the initial state flavor of the $B_{s}$ meson helps resolve the 
sign ambiguity on $\phi_{s}^{J/\psi\phi}$ for a given $\Delta \Gamma_{s}$ and improves the precision of results.
The CDF collaboration used the data sample corresponding to 1.7~fb$^{-1}$ (this result was updated recently using 2.8~fb$^{-1}$
data sample~\cite{cdf_newjpsiphi}) 
and the D\O\   data sample corresponds to 2.8~fb$^{-1}$. 
Constraints on the strong phases $\delta_{i}$ from the $B_{d}$ system allow the sign ambiguity 
on $\Delta \Gamma_{s}$ to be resolved (see recent theoretical work~\cite{gronau}). 
The CDF and D\O\   results obtained with different constraints, and compared  with the SM prediction 
are shown
in figure~\ref{fig:phis_comb}. 
The HFAG combination~\cite{hfag} gives two sets of the numerical results 
obtained with free phases $\delta_{i}$:
\begin{itemize}
\item {$\Delta \Gamma_{s} = 0.154^{+0.054}_{-0.070}$~ps$^{-1}$, \\$\phi_{s}^{J/\psi\phi}=-2\beta_{s}^{J/\psi\phi}=-0.77^{+0.29}_{-0.37}$~rad;}
\item {$\Delta \Gamma_{s} = -0.154_{-0.054}^{+0.070}$~ps$^{-1}$,\\ $\phi_{s}^{J/\psi\phi}=-2\beta_{s}^{J/\psi\phi}=2.36_{-0.29}^{+0.37}$~rad.}
\end{itemize} 

\section{$\Delta\Gamma_{s}$ from $B_{s}\to D^{(*)}_{s}D^{(*)}_{s}$~\cite{d0_delta_gamma}}
 The final state in $B_{s}\to D^{(*)}_{s}D^{(*)}_{s}$  decays is CP-even, although the theoretical uncertainty is
not well understood (see for example the discussion in~\cite{lenz}). 
The branching ratio $Br(B_{s}\to D^{(*)}_{s}D^{(*)}_{s})$ could be used to calculate the decay width difference 
from the following equation derived in~\cite{dunietz}:
\begin{eqnarray}
2\cdot Br(B_{s}\to D^{(*)}_{s}D^{(*)}_{s}) \approx \nonumber \\ 
\Delta \Gamma_{s}^{CP} \left( \frac{1+\cos\phi_{s}}{2\Gamma_{L}} + \frac{1-\cos \phi_{s}}{2\Gamma_{H}} \right). 
\label{eq:dgamma1}
\end{eqnarray}
The efficient muon triggering system at D\O\  provides access to these decays through the muons from
semileptonic decays of $D_{s}$ mesons:  
$B_{s}\to D^{(*)}_{s}(\to(\gamma)\phi\pi)D^{(*)}_{s}(\to(\gamma)\phi\mu\nu)$.
Narrow invariant mass peaks from the $\phi\to KK$ decays help to ensure a clean data sample. The number
of selected $D_{s}\phi\mu$ candidates was determined to be $N(D_{s}\phi\mu)=31.0\pm 9.4$ and the 
resulting branching ratio: $Br(B_{s}\to D^{(*)}_{s}D^{(*)}_{s})=0.042\pm 0.015(stat.)\pm 0.017(syst.)$. 
Assuming no new physics effects 
(the SM value of $\phi_{s}^{SM}\approx 0.004$) one can estimate the decay width difference from equation~\ref{eq:dgamma1}:
$\frac{\Delta\Gamma_{s}}{\Gamma_{s}}=0.088\pm0.030(stat.)\pm0.036(syst.)$.

\section{CP violation}
\subsection{$B_{s}$ semileptonic decays~\cite{cdf_dimu,d0_jpsidimu_comb}}
The SM prediction for the semileptonic asymmetry in $B_{s}$ decays is very small~\cite{lenz}:
$a_{sl}^{s}=(2.06\pm 0.57)\cdot 10^{-5}$. At the Tevatron the semileptonic asymmetry could be 
measured either from an inclusive di-muon sample, where
$ a_{sl}^{s}\sim \frac{ N_{\mu^{+}\mu^{+}} - N_{\mu^{-}\mu^{-}} }{ N_{\mu^{+}\mu^{+}}+ N_{\mu^{-}\mu^{-}} } $,
or by using the $XD_{s}\mu$ sample. The first method has very high statistical accuracy, but requires knowledge 
of asymmetries of other contributing processes in addition to the detector charge asymmetries. The 
second method has less statistical power but ensures that the major contribution to the asymmetry 
comes from the $B_{s}$ decays. The time-integrated untagged analysis at D\O\   used the data 
sample $XD_{s}(\phi\pi)\mu$ corresponding to 1.3~fb$^{-1}$ to obtain the result $a_{sl}^{s}=0.0245\pm 0.0193\pm 0.0035$~\cite{kostya}
(this analysis was recently updated using a larger data sample, time dependence and initial-state tagging~\cite{d0_newasl}).
The description of the method used to determine 
the detector asymmetries at D\O\    is given in~\cite{bruce} where the di-muon asymmetry result is presented. The combination 
of the two D\O\  results is published in~\cite{d0_jpsidimu_comb}: $a_{sl}^{s}=0.0001\pm 0.0090$. 

  The preliminary CDF result obtained using the inclusive di-muon sample is shown in~\cite{cdf_dimu}: $a_{sl}^{s}=0.020\pm 0.021\pm 0.018$.

\subsection{$B_{s}\to K\pi$~\cite{cdf_kpi}}
\begin{figure}[htb]
\vspace{9pt}
\epsfig{figure=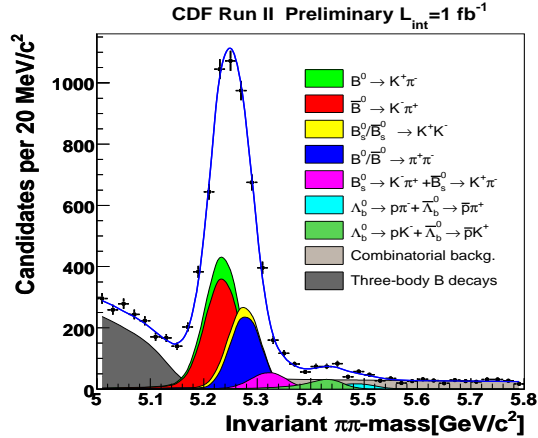,width=0.45\textwidth,height=0.3\textheight}
\caption{Invariant $\pi\pi$-mass of $B\to h^{+}h^{-}$ candidates where $h$ is $K$ or $\pi$.}
\label{fig:cdf_bskpi}
\end{figure}

  The recent measurements of direct CP violation in the $B_{u}$ and $B_{d}$ systems led to the $K\pi$ puzzle (see for example~\cite{chiang}).
It is therefore important to measure similar signatures in the $B_{s}$ system (see discussion in~\cite{fleischer}). 
 The impact parameter triggers at CDF are used to collect the hadronic $B$-decays which are necessary 
for studies of CP eigenstates. Figure~\ref{fig:cdf_bskpi} shows the invariant mass distribution for 
2 body hadronic decays at CDF assuming pion masses for the tracks. 
This sample allowed CDF to produce the only measurement of direct CP violation
in the $B_{s}$ system:
$A_{CP}^{B_{s}\to K^{-}\pi^{+}}= 
 \frac{N_{\bar{B}_{s}\to K^{+}\pi^{-}} - N_{B_{s}\to K^{-}\pi^{+}}}
{N_{\bar{B}_{s}\to K^{+}\pi^{-}} + N_{B_{s}\to K^{-}\pi^{+}}} =  
 0.39\pm 0.15(stat.) \pm 0.08(syst.). 
$
This result is in agreement with the theoretical predictions~\cite{chiang1}.

\subsection{$B^{\pm}\to D^{0} K^{\pm}$~\cite{cdf_dk}}
CP analysis of the decays
 $B^{\pm}\to D^{0}K^{\pm}$ represents another example of using the impact parameter triggers at CDF. The CP asymmetry
is measured in these decays.
This asymmetry is relevant for the determination of the CKM angle $\gamma$ (see~\cite{gronau1,gronau2}). The $D^{0}$ meson
decays either to CP eigenstates $K^{+}K^{-}\ (\pi^{+}\pi^{-})$, or flavour eigenstate $K^{-}\pi^{+}$.
The different modes were separated using invariant masses, momentum imbalance, total momentum and $dE/dx$ information. 
The measured asymmetries agree with the B-factory measurements~\cite{d0k_babar,d0k_belle} and have competitive precision:
$A_{CP+}=\frac{Br(B^{-}\to D^{0}_{CP+}K^{-}) - Br(B^{+}\to D^{0}_{CP+}K^{+})}
{Br(B^{-}\to D^{0}_{CP+}K^{-}) + Br(B^{+}\to D^{0}_{CP+}K^{+})}=0.37\pm 0.14(stat.)\pm 0.04(syst.)$,
 $R=\frac{Br(B^{-}\to D^{0}K^{-}) - Br(B^{+}\to \bar{D}^{0}K^{+})}
{Br(B^{-}\to D^{0}K^{-}) + Br(B^{+}\to \bar{D}^{0}K^{+})}=0.0745\pm 0.0043(stat.)\pm 0.0045(syst.)$,
$R_{CP+}=\frac{Br(B^{-}\to D^{0}_{CP+}K^{-}) + Br(B^{+}\to D^{0}_{CP+}K^{+})}
{(Br(B^{-}\to D^{0}K^{-}) + Br(B^{+}\to \bar{D}^{0}K^{+}))/2}=1.57\pm 0.24(stat.)\pm 0.12(syst.)$.

\subsection{$B^{\pm}\to J/\psi K^{\pm}$~\cite{d0_jpsik}}
Highly efficient di-muon triggers allow for collection of large $J/\psi$ samples at Tevatron which are competitive 
with the B-factories in some of the CP violation measurements in $B_{u}$ and $B_{d}$ systems. 
The SM expectation of the direct CP asymmetry in $B^{\pm}\to J/\psi K^{\pm}$ decays is small: $A_{SM}^{J/\psi K} \sim 0.003$~\cite{hou}.
However, the new physics effects (for example a fourth generation) may enhance the asymmetry up to 0.01. The D\O\   experiment 
measured this asymmetry 
to be $A^{J/\psi K}=+0.0075 \pm 0.0061(stat.)\pm 0.0027(syst.)$.

\section{Conclusion}
So far, the Tevatron experiments CDF and D\O\  have provided the unique sources of the CP parameters in $B_{s}$ system. 
The complimentary measurements in $B_{u/d}$  systems are not only supporting the $B_{s}$ analyses but also 
providing results competitive with the B-factories. Interestingly though, the CP parameters 
in the $B_{s}$  system start showing some deviations from the SM predictions in addition to the 
$B_{u/d}$ puzzles. All analyses are statistically limited and the precision will improve in the near future.

  The author thanks the organisers of the QCD08 Conference for a very interesting program and the physics
analysis representatives from CDF and D\O\   for providing results.

\end{document}